\documentclass[prl,showpacs,twocolumn,amsmath,amssymb]{revtex4}
\usepackage{bm}
\usepackage{graphicx}
\usepackage{bbold}

\newcommand{\pdagger}{{\phantom{\dagger}}}
\newcommand{\dt}{\Delta\tau}
\newcommand{\refq}[1]{(\ref{eq:#1})}
\newcommand{\reff}[1]{Fig.\ \ref{fig:#1}}

\begin{document}

\title{Reply to ``Comment on `Orbital-selective Mott transitions in the\\ anisotropic two-band Hubbard model at
finite temperatures' ''}

\author{C.~Knecht}
\author{N.~Bl\"umer}
\email{Nils.Bluemer@uni-mainz.de}
\author{P.\ G.\ J.\ van Dongen}
\affiliation{Institute of Physics, Johannes Gutenberg University, 55099 Mainz, Germany}

\date{\today}

  \begin{abstract}
    In a Comment [cond-mat/0506138] on our recent e-print [cond-mat/0505106]
    Liebsch claimed ``excellent correspondence''
    between our high-precision quantum Monte-Carlo (QMC) data for the
    anisotropic two-band Hubbard model with Ising type exchange
    couplings and his earlier QMC results. Liebsch
    also claimed that the sequence of two orbital-selective Mott
    transitions, identified by us in this model, had already been
    reported in his earlier work.
    Here we demonstrate that both claims are incorrect. We establish
    that Liebsch's previous QMC estimates for the quasiparticle weight
    $Z$ have relative errors exceeding $100\%$ near transitions and
    cannot be used to infer the existence of a second Mott transition
    (for $U_{c2}\approx 2.5$). We further show that Liebsch's attribution
    of our findings to his own earlier work is disproved by the
    published record.  Consequently, the Comment is
    unwarranted; all results and formulations of our e-print remain valid.

  \end{abstract}
  \maketitle

  In a recent e-print \cite{Knecht05} we studied the anisotropic degenerate
  two-orbital Hubbard model (notation as in \cite{Knecht05})
  \begin{eqnarray}
  H&=&-\sum_{\langle ij\rangle m\sigma}  t^\pdagger_m c^{\dag}_{im\sigma}
  c^\pdagger_{jm\sigma}\,+\,U\sum_{im}n_{im\uparrow} n_{im\downarrow}\nonumber\\
    &&+\sum\nolimits_{i\sigma\sigma'}(U'-\delta^\pdagger_{\sigma \sigma'} J_z)n^\pdagger_{i1\sigma}
  n^\pdagger_{i2\sigma'}\label{eq:Hamiltonian}
  \end{eqnarray}

\vspace{-0.5em}\noindent
  using high-precision quantum Monte Carlo (QMC) simulations and
  showed (in accordance with independent slave-spin calculations
  \cite{deMedici05}) that this model contains \emph{two} consecutive
  metal-insulator transitions and, hence, describes an
  ``orbital-selective Mott transition'' (OSMT) \cite{Anisimov02}.  This finding is to be contrasted with earlier
  results by
  Liebsch \cite{Liebsch03a,Liebsch03b,Liebsch04} who reported
  a single Mott transition and explicitly excluded the occurrence of
  an OSMT for the same model. Our correction of this earlier picture established
  \refq{Hamiltonian} as a {\em minimal model} for the OSMT, seen
  experimentally in Ca$_{2-x}$Sr$_{x}$RuO$_4$, \cite{Nakatsuji00ab} and demonstrated
  that the inclusion of spin-flip and pair-hopping terms (see
  \cite{Knecht05} and references therein) is not essential in this
  respect.

  While the high accuracy of our QMC data \cite{Knecht05} and the validity of our
  conclusions have not been challenged in any publication or
  e-print, Liebsch took exception to our discussion of his
  earlier work, first in \cite{Liebsch05a} and then, in the form of a
  Comment on our work, in \cite{Liebsch05b}. Liebsch now claims
  \cite{Liebsch05a, Liebsch05b} that he not only observed the second
  transition but also reported on it in his earlier work
  \cite{Liebsch03b,Liebsch04}, and that, when he systematically
  mentioned only one single transition, he in fact meant two (one
  first-order and one second-order transition).  Liebsch also claims
  \cite{Liebsch05b} that the numerical data in our e-print
  \cite{Knecht05} are in ``excellent correspondence'' with his earlier
  work \cite{Liebsch04}, the implication being that any transition
  seen by us must necessarily also be contained in Liebsch's earlier
  work.
  
  The purpose of this Reply to Liebsch's Comment \cite{Liebsch05b} is
  twofold.  First we show by a detailed comparison of both sets of QMC
  results that the data of \cite{Liebsch04} are of insufficient
  quality near $U_{c2}$ and, hence, do not allow for any statement
  concerning the occurrence or non-occurrence of a second transition.
  In this crucial interaction regime near $U_{c2}$ there is,
  therefore, \emph{no} ``excellent correspondence'' between both data
  sets.  Secondly, we demonstrate on the basis of explicit quotations
  from the published record that Liebsch did \emph{not} identify a
  second transition in his earlier work. In fact he is generally
  viewed as the proponent of the single-transition scenario for this
  model and publicly attempted to adjust \cite{Liebsch05a, Liebsch05b}
  this view only \emph{after} submission of our paper \cite{Knecht05}.
  In addition to these two main issues, we comment on the quality of
  Liebsch's QMC data near the first transition at $U_{c1}$ and his
  quasiparticle spectra, and we detail our reanalysis \cite{Knecht05}
  of Liebsch's earlier IPT results~\cite{Liebsch04}.

  
  {\it Comparison of QMC data --} In the following, we compare our QMC
  estimates for the orbital-dependent quasiparticle weights
  $Z_{\text{wide}}$, $Z_{\text{narrow}}$ \cite{Knecht05} with
  corresponding data by Liebsch \cite{Liebsch04} as well as with new
  numerically exact QMC data \cite{note1}. We show that (i)~our
  published QMC data is extremely accurate, with typical relative
  errors of $10^{-2}$, (ii)~a careful analysis of this data reveals a
  second Mott transition at $U_{c2}\approx 2.5$, and (iii)~Liebsch's
  QMC data is much too noisy in the OSMT region -- with relative
  errors exceeding $100\%$ near {\em both} transitions -- to detect
  the second transition.
  In \reff{Zcomp}a,
  \begin{figure}
  \includegraphics[width=\columnwidth,clip=true]{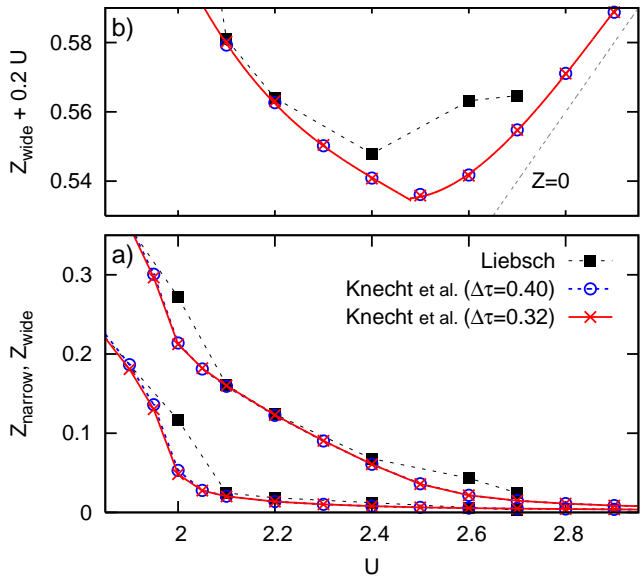}
  \caption{Comparison of discrete quasiparticle weights $Z$ at $T=1/32$:
    a) high-precision QMC data with minimal $\dt$ dependence
    \cite{Knecht05} (circles, crosses) clearly shows kinks in $Z$ for
    narrow/wide band (lower/upper set of curves and symbols) at
    $U_{c1}\approx 2.0$; Liebsch's QMC data \cite{Liebsch04} (squares)
    deviates markedly near $U_{c1}$ and for $U\approx 2.6$. b) A
    subtle kink in the QMC data of Ref.\ \cite{Knecht05} -- clearly
    visible only after adding a linear term -- indicates a second
    transition at $U_{c2}\approx 2.5$; this signal is lost
    in the noise of the data of Ref.\ \cite{Liebsch04}.}
  \label{fig:Zcomp}
  \end{figure}
  the lower/upper sets of symbols and curves correspond to
  quasiparticle weights for the narrow/wide band.  Our QMC results
  for discretizations $\dt=0.4$ and $\dt=0.32$ (circles/crosses) are
  practically on top of each other; a first kink in these estimates
  for both $Z_{\text{wide}}$ and $Z_{\text{narrow}}$ clearly signals a
  transition at $U_{c1}\approx 2.0$. Even on this scale (much larger
  than in \cite{Knecht05}, \cite{Liebsch05b}), the significance of a
  change of slope in $Z_{\text{wide}}$ in the region $2.4\le U\le 2.7$
  is unclear. Only the analysis shown in \reff{Zcomp}b using a linear
  offset (as used for Figs.\ 1 and 5 in \cite{Knecht05}) reveals a
  well-localized kink and, thus, a second transition at $U_{c2}\approx
  2.5$.
  Now consider Liebsch's QMC data (squares in~\reff{Zcomp}): already
  on the scale of \reff{Zcomp}a, it deviates strongly from our
  high-precision results near both transition regions. In particular,
  it cannot be used to detect the subtle transition at $U_{c2}$, as
  demonstrated in \reff{Zcomp}b: all changes in slope of Liebsch's
  estimates for $Z_{\text{wide}}$ (squares and dashed line) are
  dominated by numerical errors; this data even displays an unphysical
  negative curvature of $Z_{\text{wide}}$ near $U=2.6$. Liebsch's data
  also suffers from too coarse a grid and, in particular, from the
  lack of data points beyond the retroactively claimed
  \cite{Liebsch05b} transition. We further stress that Liebsch's criterion that
  $Z_{\text{wide}}$ ``becomes very small near $U_{b}\approx 2.7$ eV''
  \cite{Liebsch05b} is not a valid criterion for a transition.

  In order to fully quantify the numerical errors (including Monte
  Carlo statistical errors, convergence errors and discretization
  errors) of all data sets shown above, we have extended our
  QMC simulations \cite{Knecht05} to smaller discretizations
  ($\dt=0.25$ and $\dt=0.20$) and have obtained numerically exact
  reference data by explicit extra\-polations $\dt\to 0$. Relative
  errors computed by comparison with this reference data are shown
  in \reff{Zdevs}
  \begin{figure}
  \includegraphics[width=\columnwidth,clip=true]{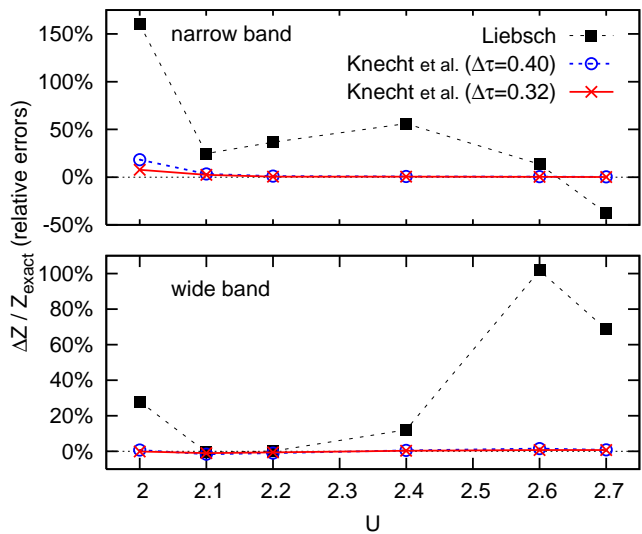}
  \caption{Estimates of relative errors in $Z$ at $T=1/32$: Liebsch's
    QMC results (squares) deviate significantly from new numerically
    exact QMC data -- by more than 100\% near transitions.  In
    contrast, already our \cite{Knecht05} finite-$\dt$ QMC data
    (circles, crosses) is accurate within about $1\%$ (except for
    $Z_{\text{narrow}}$ at $U=2.0$).}
  \label{fig:Zdevs}
  \end{figure}
  for both the wide and narrow band (lower and upper panel,
  respectively). The relative errors in our QMC data
  (circles, crosses) are visible only at the narrow-band Mott
  transition (in $Z_{\text{narrow}}$ for $U=2.0$); throughout the rest
  of the OSMT region, our relative errors are typically smaller than
  $10^{-2}$.  The errors of Liebsch's data (squares in~\reff{Zdevs})
  are in this region two orders of magnitude larger and exceed $100\%$
  at both transitions (as identified by us); only in the metallic
  phase for $U\le 1.6$ (not included in this comparison), Liebsch's
  data appear reasonably accurate. We conclude that the second transition, identified by us,
  cannot be seen in Liebsch's QMC estimates of $Z$. This
  already disqualifies the Comment.


  {\it Liebsch's published record on OSMT --} We now come to the
  second part of our Reply, in which we show on the basis of the
  published record, that Liebsch \cite{Liebsch03a, Liebsch03b,
    Liebsch04} in fact proposed a scenario very different from the
  OSMT scenario found by us, namely the existence of only a single
  Mott transition.  Indeed, up to the recent surge of e-prints on the
  subject, \cite{Koga05, Ferrero05, deMedici05, Arita05, Knecht05}
  Liebsch was generally recognized as the proponent of the
  single-transition scenario, as witnessed, e.g., by various recent
  statements to this effect in Refs.\ \cite{Koga04b, Ferrero05,
    deMedici05, Arita05}.  It was only {\em after} submission of our
  paper that Liebsch \cite{Liebsch05a,Liebsch05b} attempted to
  reinterpret his earlier work \cite{Liebsch03a,Liebsch03b,Liebsch04}.
  That such a reinterpretation is in fact precluded by the published
  record will now be shown.
  
  Chronologically Liebsch's first paper on the anisotropic degenerate
  multi-orbital Hubbard model is \cite{Liebsch03a}, which carried the
  title \textsl{``Absence of orbital-dependent Mott transition in
    Ca$_{2-x}$Sr$_x$RuO$_4$''}. The abstract informs us, that the
  results suggest ``a common metal-insulator transition for all
  $t_{2g}$ bands at the same critical $U$''. Liebsch explains on the
  basis of general arguments from the theory of phase transitions (his
  Ref.\ 17) that in several ``conceptually closely related'' problems
  ``a true separation of phases does not seem possible''. In
  accordance with this expectation, Liebsch then finds, using
  QMC/DMFT, ``a behavior that differs fundamentally from the one
  reported'' by Anisimov et al.\ \cite{Anisimov02}, characterized by
  ``one common metal-insulator transition [existing] at an
  intermediate critical value of $U$ between those of the isolated
  $d_{xz,yz}$ and $d_{xy}$ bands''. Needless to say that this finding
  is irreconcilable with the existence of two separate transitions. In
  \cite{Liebsch03a}, Liebsch calculates the density of states of
  Sr$_2$RuO$_4$ for various parameter values and temperatures and
  deduces from the results the existence of ``a common Mott
  transition'' satisfying ``$U^{\rm crit}_{xz,yz}<U^{\rm
    crit}_{t2g}<U^{\rm crit}_{xy}$, i.e., the Mott transition for the
  actual $t_{2g}$ complex requires a critical $U$ between those of the
  independent $d_{xz,yz}$ and $d_{xy}$ bands''. Again, this leaves no
  room for two separate Mott transitions.

  Chronologically next in line is Ref.\ \cite{Liebsch03b}, the
  abstract of which informs us, that ``interorbital Coulomb
  interactions in nonisotropic multiorbital materials give rise to a
  single Mott transition'', implying (according to the third paragraph
  of the text) ``all subbands to be either metallic or insulating''.
  The text then goes on: ``The critical Coulomb energy $U_c$ of the
  interacting system lies between the $U_{ci}$ of the isolated
  subbands. Thus, narrow subbands are less correlated and wide
  subbands are more correlated than in the absence of interorbital
  Coulomb interactions. Coexistence of metallic and insulating
  behavior in different subbands does not occur.''  Similar statements
  are repeated both in the bulk of the text and in the captions to
  Figs.\ 1 and 2 of that paper, leading to the conclusion that ``the
  present picture does not support the `orbital-selective' Mott
  transitions''.

  Last in line of the early works of Liebsch on this subject is Ref.\
  \cite{Liebsch04}, which has as its title ``Single Mott transition in
  the multiorbital Hubbard model'', while the abstract states that
  both the IPT approximation and QMC ``give rise to a single
  first-order transition rather than a sequence of orbital-selective
  transitions''. This conclusion of Ref.\ \cite{Liebsch04}, in
  combination with the alleged ``complex'' finite temperature behavior
  of the subbands, constitutes a serious misinterpretation
  of the orbital-selective Mott transition, which actually occurs in
  this model. Ref.\ \cite{Liebsch04} states (in Section II) that the
  quasiparticle weights obtained from IPT ``exhibit
  first-order transitions at precisely the same critical $U_c(T)$.
  This picture supports our previous results suggesting a single Mott
  transition in a nonisotropic multiorbital environment, in contrast
  to the orbital selective transitions found [by Anisimov et al.\ and
  Koga et al.].'' We shall see below that in fact
  IPT, too, predicts an orbital-selective Mott transition. Using the QMC-DMFT, Liebsch then shows (in Section
  III) that, ``as in the case of IPT, there is no evidence for a
  second transition in the wide subband at larger $U$''. The paper
  concludes (in Section IV), that ``the present results confirm our
  previous finding, namely, the existence of a single transition
  rather than a sequence of orbital-selective transitions as the
  on-site Coulomb energy is increased''. Thus the occurrence of an
  OSMT is explicitly denied in both Ref.\ \cite{Liebsch03b} and Ref.\ \cite{Liebsch04}.

  We now consider Liebsch's Comment \cite{Liebsch05b}. In the Comment,
  Liebsch ignores his previous work
  \cite{Liebsch03a,Liebsch03b} and focuses on his QMC calculations at
  $T=0.031$ eV in \cite{Liebsch04}. This focus on Ref.\
  \cite{Liebsch04} is remarkable, since Ref.\ \cite{Liebsch04}, as we
  saw above, just confirms his ``previous finding''. In
  \cite{Liebsch05b} it is then stated that in \cite{Liebsch04} two
  transition regions (at $U_a$ and $U_b$) were identified, with
  bad-metal behavior and a breakdown of Fermi-liquid behavior for the
  wide band in the ``intermediate phase'' $U_a<U<U_b$. Here we wish to
  record that in \cite{Liebsch04} neither the notation $U_a$ nor $U_b$ was used, that
  ``Fermi-liquid behavior'' (or the breakdown thereof) was not
  mentioned in the entire paper, that Liebsch in fact argued (by
  comparison with data for $T=0.02$ eV) that ``both subbands undergo a
  common transition at the same Coulomb energy and that the wide
  subband exhibits pronounced bad-metal behavior above $U_c$'' and
  that he concluded (see also above) that ``there is no evidence for a
  second transition in the wide subband at larger $U$''. We further
  note that, since there is no ``second transition'' and no ``$U_b$''
  in \cite{Liebsch04}, the hysteresis behavior cannot have suggested
  that ``only the lower transition is first-order''. In
  \cite{Liebsch05b} Liebsch states that he argued in \cite{Liebsch04}
  ``that the Mott transition in the nonisotropic Hubbard model is
  governed by a \emph{single first-order transition} \dots'',
  suggesting that he left room for a second continuous transition at
  some larger value of $U$. Comparison with the above quotations
  from the published record shows that this interpretation
  of Ref. \cite{Liebsch04} is untenable.
  
  We briefly comment on Ref. \cite{Liebsch05a}, in which our work
  \cite{Knecht05} is also cited. While \cite{Liebsch05a} is, generally
  speaking, an interesting extension of earlier work \cite{Koga04a} to
  finite temperatures, this preprint, too, contains several
  inaccuracies.  For instance, it is \emph{not} true that (as claimed
  in the second column) the phase diagram of \cite{Liebsch05a} is in
  agreement with that of \cite{Liebsch04} (cf.\ Fig.\ 5 in
  \cite{Liebsch04} and Fig.\ 4 in \cite{Liebsch05a}), or that (fourth
  column) the two-transition scenario of Fig. 2 ``confirms the
  picture'' of Ref.\ \cite{Liebsch04}, or that (sixth column)
  Liebsch's use of ``single Mott transition'' \cite{Liebsch04} did not
  ``imply the non-existence of the continuous transition \dots ''. All
  of this is irreconcilable with the passages from the published
  record, quoted above. We further note that ``breakdown of
  Fermi-liquid behavior'' occurs in this paper for the first time, not
  in \cite{Liebsch04} (as claimed in \cite{Liebsch05b}).

  {\it Comment on Liebsch's spectra --} Quasiparticle 
  spectra play a subordinate role in Liebsch's work
  \cite{Liebsch03a,Liebsch03b,Liebsch04}, according to Ref.\ 
  \cite{Liebsch04} in order to ``avoid the uncertainties from the
  maximum entropy reconstruction''. Here we discuss them nonetheless,
  primarily because our spectra \cite{Knecht05} are compared to those
  of Ref.\ \cite{Liebsch04} in \cite{Liebsch05b}.  Generally,
  Liebsch's spectra (obtained from QMC using the maximum entropy
  method) appear overly broad with strong high-frequency tails and
  very smooth behavior at low frequencies. This lack of resolution
  obscures the observation of gaps: e.g., the narrow-band spectrum for
  $U=2.1$ (solid line in upper-most panel of Fig.\ 2 in
  \cite{Liebsch05b}) appears as still metallic [with
  $N(\omega\!=\!0)\approx 0.022$, more than $3\%$ of the
  noninteracting value]; a true gap for this interaction [with
  $N(\omega\!=\!0)=0$] has only been established in our work (cf.\ 
  Fig.\ 4 in \cite{Knecht05}). In Liebsch's data, this gap in the
  narrow band is only well-established at $U=2.4$ (second panel of
  Fig.\ 2 in \cite{Liebsch05b}). The lack of a visible gap in the
  wide-band spectrum is attributed in \cite{Liebsch05b} to the filling
  of a pre-existing gap at finite temperatures: ``Only if $U$ is
  increased to about $2.7$ eV is the gap in the wide band large enough
  to not be {\em obliterated} by this temperature broadening.''  The
  delayed opening of gaps (with respect to the ``single Mott
  transition'' pinpointed at $U=2.1$) is explicitly treated on equal
  footing for both bands: ``For $U=2.7$ eV \dots a gap opens up [in
  the wide band].
  \begin{figure}[t]
  \includegraphics[width=\columnwidth,clip=true]{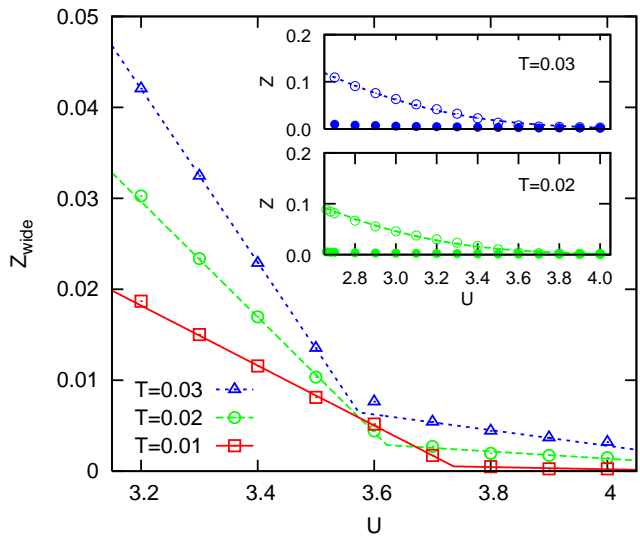}
  \caption{Reanalysis of IPT estimates for $Z$: A high-resolution plot
    of $Z_{\text{wide}}$ (extracted from the preprint to
    \cite{Liebsch04}) reveals transitions near $U_{c2}\approx 3.6-3.7$
    as kinks in otherwise linear fits which represent the data very
    well for $T\le 0.02$. The rounding-off at $T=0.03$ (and $U=2.6$)
    -- if genuine -- may indicate that $T^*_{\text{IPT}}<0.03$.
    Insets: the transitions apparent in the main panel are obscured in
    representations similar to \cite{Liebsch04,Liebsch05b} -- on this
    scale, the IPT data is consistent with featureless fits (e.g., of the
  form $a \exp[-b U^c]$ -- dashed lines).}
  \label{fig:Z_IPT}
  \end{figure}
  The narrow band undergoes a {\em similar cross-over} behavior,
  except at slightly lower values of $U$.''  \cite{Liebsch05b} We
  conclude that the broadness of Liebsch's spectra (which must be
  attributed to noisy QMC data) prevents a proper characterization of
  phases as metallic or insulating; in addition, more grid points
  would have been necessary in order to see the second transition.  In
  comparison, our spectra show well-defined gaps with sharp edges
  (Fig.\ 3 in \cite{Knecht05} and lower three panels of Fig.\ 2 in
  \cite{Liebsch05b}); still, we would not have relied on spectra alone
  for the identification of the second transition.

  
  {\it Reanalysis of IPT-data of Ref.\ \cite{Liebsch04} --} Finally we
  prove our previous statement that, in contrast to Liebsch's findings
  in \cite{Liebsch03b,Liebsch04}, the IPT approximation, too,
  describes \emph{two} separate transitions rather than ``a single
  Mott transition'' \cite{Liebsch04}. For this purpose we use
  Liebsch's own IPT-data \cite{Liebsch04} for the quasi-particle
  weight, which is plotted in \reff{Z_IPT}.  The insets show data for
  $T=0.02$ and $T=0.03$ for both the narrow and the wide band. The
  main panel shows $Z_{\rm wide}(U)$ for various temperatures and
  $3.2\leq U\leq 4.0$. As can be seen from \reff{Z_IPT}, the behavior
  of $Z_{\rm wide}(U)$ is approximately piecewise linear on both sides
  of $U_{c2}(T)\simeq 3.6-3.7$, where $U_{c2}(T)$ increases slightly
  with decreasing temperature. This second transition at $U_{c2}$ is
  not visible in the small-scale graphical representation of the IPT
  data in Figs.\ 1-4 of [7] (similar to the insets in \reff{Z_IPT})
  and was mentioned by Liebsch \cite{Liebsch05b} only \emph{after} he
  was informed of its existence by us \cite{note:IPT}.

  
  {\it Summary --} We have shown that, in contrast to claims in
  \cite{Liebsch05b}, there is \emph{no} ``excellent correspondence''
  between the numerical data of our paper \cite{Knecht05} and those of
  Ref.\ \cite{Liebsch04}, and that the conclusions reached in
  \cite{Knecht05} indeed correct previous erroneous statements
  concerning ``single Mott transitions'' in \cite{Liebsch04}. The
  published record clearly demonstrates that Liebsch
  \cite{Liebsch03a,Liebsch03b,Liebsch04} did \emph{not} observe two
  separate transitions. A comparison of Liebsch's QMC data with our
  high-precision results further shows that the second transition was
  hidden in \cite{Liebsch04} by huge error bars (of up to 100\% in
  the critical region) and inadequate data analysis.
  
  We thank E.\ Jeckelmann and D.\ Vollhardt for discussions and
  acknowledge support by the NIC J\"{u}lich and by the DFG
  (Forschergruppe 559, Bl775/1). 

\vspace{-0.4cm}

\begin{thebibliography}{16}
\bibitem{Knecht05} C.\ Knecht, N.\ Bl\"umer and P.\ G.\ J.\ van Dongen, cond-mat/0505106 (unpublished).
\bibitem{deMedici05} L.~de' Medici, A.\ Georges, and S.\ Biermann, cond-mat/0504040 (unpublished).
\bibitem{Anisimov02} V.I.\ Anisimov, I.A.\ Nekrasov, D.E.\ Kondakov, T.M.\ Rice, and M. Sigrist, Eur.\ Phys.\ J. B {\bf 25}, 191 (2002).
\bibitem{Liebsch03a} A.\ Liebsch, Europhysics Letters {\bf 63}, 97 (2003).
\bibitem{Liebsch03b} A.\ Liebsch, Phys.\ Rev.\ Lett.\ {\bf 91}, 226401 (2003).
\bibitem{Liebsch04} A.\ Liebsch, Phys.\ Rev.\ B {\bf 70}, 165103 (2004).
\bibitem{Nakatsuji00ab} S.\ Nakatsuji, Y. Maeno, Phys.\ Rev.\ Lett.\ {\bf 84}, 2666 (2000); Phys.\ Rev.\ B {\bf 62}, 6458 (2000).
\bibitem{Liebsch05a} A.\ Liebsch, cond-mat/0505393 (unpublished).
\bibitem{Liebsch05b} A.\ Liebsch, cond-mat/0506138 (unpublished).
\bibitem{note1} Note that $Z_{\text{wide}}$ and $Z_{\text{narrow}}$ are quite insensitive to OSMTs
(identified in \cite{Knecht05} using more sophisticated criteria);
  thus, only very precise data has significance with respect to
  existence and position of a second transition.
\bibitem{Koga05} A.\ Koga, N.\ Kawakami, T.M.\ Rice, and M.\
Sigrist, cond-mat/0503651 (unpublished).
\bibitem{Ferrero05} M.\ Ferrero, F.\ Becca, M.\ Fabrizio, and M.\ Capone, cond-mat/0503759 (unpublished).
\bibitem{Arita05} R.\ Arita and K.\ Held, cond-mat/0503764 (unpublished).
\bibitem{Koga04b} A.\ Koga, N.\ Kawakami, T.\ Rice, and M.\ Sigrist, cond-mat/0406457 (unpublished).
\bibitem{Koga04a} A.\ Koga, N.\ Kawakami, T.M.\ Rice, and M.\ Sigrist, Phys.\ Rev.\ Lett. {\bf 92}, 216402 (2004).
\bibitem{note:IPT} Private communication of N.  Bl\"umer to A. Liebsch
  on May 4th, 2005.
\end{thebibliography}
\end{document}